\documentclass[12pt]{article}
\usepackage{amssymb}
\usepackage{epsfig}

\newcommand{\bq}{\begin{equation}}
\newcommand{\eq}{\end{equation}}
\newcommand{\bqa}{\begin{eqnarray}}
\newcommand{\eqa}{\end{eqnarray}}
\newcommand{\ben}{\begin{enumerate}}
\newcommand{\een}{\end{enumerate}}
\newcommand{\bc}{\begin{center}}
\newcommand{\ec}{\end{center}}

\def\lsim{\lesssim}

\def\pr#1#2#3{ Phys. Rev. ${\bf{#1}}$ (#2) #3}
\def\prl#1#2#3{ Phys. Rev. Lett. ${\bf{#1}}$ (#2) #3}
\def\pl#1#2#3{ Phys. Lett. ${\bf{#1}}$ (#2) #3}

\def\np#1#2#3{ Nucl. Phys. ${\bf{#1}}$ (#2) #3}
\def\zp#1#2#3{ Z. f. Phys. ${\bf{#1}}$ (#2) #3}

\def\etal{{\it et.al.\/}}

\global\nulldelimiterspace = 0pt

\def\O{ {\cal O }}

\def\swd{s^2_W}
\def\cwd{c^2_W}

\def\mw{M_W}

\def\mh{m_H}

\begin{document}
\thispagestyle{empty}
\begin {flushleft}

PM/97-28\\
THES-TP 97/06\\
August 1997\\
\end{flushleft}

\vspace*{2cm}

\hspace*{-0.5cm}
\begin{center}
{\Large {\bf Tests of possible non standard properties of the
heavy quarks and the Higgs boson}}
\footnote{{\bf Contribution to the studies on Physics and Detectors
for the Linear Collider"}, to appear in DESY 97-123E. 
Partially supported by the EC contract CHRX-CT94-0579.}
\hspace{2.2cm}\null 

\hspace*{-0.5cm}
\vspace{1.cm} \\{\bf \large G.J. Gounaris$^a$, J. Layssac$^b$,
D.T. Papadamou$^a$, G. Tsirigoti$^a$ and F.M.
Renard$^b$}\hspace{2.2cm}\null \\
 \vspace{0.2cm}
$^a$Department of Theoretical Physics, University of Thessaloniki,
\hspace{2.2cm}\null \\
Gr-54006, Thessaloniki, Greece.\\ \vspace{0.2cm} $^b$ Physique
Math\'{e}matique et Th\'{e}orique, UPRES-A 5032\hspace{2.2cm}\null\\
Universit\'{e} Montpellier
II,  F-34095 Montpellier Cedex 5.\hspace{2.2cm}\null\\[1cm]

\vspace*{0.1cm}
\end{center}

\vspace*{2cm}
\begin{center}
{\bf Abstract}\hspace{2.2cm}\null
\end{center}
%
%
We discuss the sensitivity of the  processes 
$e^+e^-\to b\bar b$, $e^+e^-\to~t\bar t$ and  $\gamma\gamma
\to H$  to certain residual New Physics interactions
affecting the heavy quark and Higgs boson sector. With a linear 
collider of $500 ~GeV$,
one should be able to detect and identify such effects with a
characteristic scale up to $50 ~TeV$.

\setcounter{footnote}{0} 
\clearpage
\newpage 
  
\hoffset=-1.46truecm
\voffset=-2.8truecm
\textwidth 16cm
\textheight 22cm
\setlength{\topmargin}{1.5cm}

\section{Introduction} 

We assume that a certain dynamics generically called New Physics
(NP) exists beyond the Standard Model (SM) and is responsible for many
of its unexplained features like the structure of the scalar
sector and the mass spectrum of the leptons, the
quarks and the gauge bosons. It is further assumed that this new
dynamics involves heavy degrees
of freedom characterized by an effective scale much higher than 
the electroweak scale; i.e. $\Lambda \gg \mw$. It is then
natural to expect that this dynamics which leads to the mass of the
usual particles, also generates some additional new interactions
among these particles, called residual NP effects. 
This picture suggests that the NP effects should most probably
affect particles especially concerned by the
mass generation mechanism, namely  the heavy quarks 
and the $W_L$, $Z_L$ and  scalar states. 
Inevitably, such NP will also involve "anomalous" gauge interactions,
which are required by the gauge principle 
whenever derivative interactions appear.\par

Recent hints for anomalous effects have appeared in $Z\to b\bar b$;
i.e. $\Gamma(Z\to b\bar b)$ is 1.8 $\sigma$ too high and $A_b$ is
2$\sigma$ too low, as compared to SM predictions, \cite{databb}. 
There also exist
unexplained features in the high $p_T$ jets produced at 
Tevatron \cite{dijet} and in high
$q^2$ events at HERA \cite{HERA}.\par

To describe these residual effects in a most model-independent way, we
 use the effective lagrangian method.
Thus, for $M_W\lsim E \ll \Lambda$, by integrating out
all NP degrees of freedom, one obtains an effective Lagrangian
describing the residual effects among usual particles. This Lagrangian
satisfies of course $SU(3)\times SU(2)\times U(1)$ 
gauge invariance broken at the electroweak scale, and involves a
Higgs particle which should also be of the order of the electroweak
scale or heavier. 
For   sufficiently large $\Lambda$, the lowest dimension 
operators ($d=6$) should  dominate NP contribution. 
One is then able to draw
a list of operators \cite{Buch}, 
$\O_i$, being
associated to corresponding ~dimension ($4-d=-2$) couplings $f_i$. 
These couplings are the free
parameters of the description and, through unitarity
considerations, they can be related to the NP
scale where either new strong interactions are generated
or new particles are produced \cite{unit,Vlachos}.  \par

The general description of NP in terms of the list  $\O_i$ 
involves various classes of bosonic and
fermionic operators with specific features. 
Bosonic operators are usually separated
into "non-blind", "blind" and even "super-blind" ones, depending on
their appearance in Z-peak observables at tree or loop levels,
\cite{DeR,Hag,GGbos}.
The same concept has been extended to the list of heavy quark
operators; i.e. operators involving  quarks of the third
family. In classifying them, 
the possible presence of the $t_R$ field has been used
\cite{Zbbtop,Kur}. To reduce the number of the  independent
operators, the equations of motion for 3rd-family quarks and 
scalar fields have been used, since they do
not mix other families, to the extent that all fermion masses
except the top are neglected. Thus finally we end up with three
classes of such operators involving quarks of the third family
\cite{Papa}. \par

Tests of purely bosonic operators in $e^+e^-$ annihilation
processes were discussed in previous reports,
(for a review see \cite{LEP2,LC}). In Section 2 we
report on the tests concerning the heavy quark
sector, and  Section 3 is
devoted to an a example of tests concerning operators
inducing CP violating $H \gamma \gamma$ interactions.\par

\section{Tests in the heavy quark sector}

\subsection{Tests in $e^+e^-\to t\bar t$}\par

Obviously, the heavy top should be a privileged place for looking for
these effects; so we start with the process $e^+e^-\to t\bar t$
observable at LC \cite {LC}. The aforementioned operators 
lead to modifications of the
$\gamma t\bar t$ and $Zt\bar t$ couplings. 
The general form of such
CP-conserving vertices is given by \cite{whisnant,Yuan,dj,Kur}
\begin{equation}  
\label{eq:dgz}
-i \epsilon_\mu^V J^{\mu}_V = -i e_V \epsilon_\mu^V \bar
u_t(p)[\gamma^{\mu}d^V_1(q^2)+\gamma^{\mu}\gamma^5d^V_2(q^2)
+(p-p^{\prime})^{\mu}d^V_3(q^2)/m_t] v_{\bar t}(p^{\prime}) \ ,
\end{equation}
where $\epsilon_\mu^V$ is the polarization of the vector boson
$V=\gamma, Z$. The
outgoing momenta $(p,~p^{\prime})$ refer to $(t,~\bar t) $ 
respectively and
satisfy $q\equiv p+p^{\prime}$. The normalizations are
determined by $e_{\gamma}\equiv e$ and $e_Z\equiv e/(2s_Wc_W)$,
while $d^V_i$ are in general $q^2$ dependent form factors.
At tree level SM feeds the first set, called set (1), 
of the three couplings 
\begin{equation} 
\label{eq:dgzSM}
d^{\gamma, SM0}_1 = {\frac{2}{3}} \ \ , \ \ d^{Z, SM0}_1=
g_{Vt}={\frac{1}{2}}%
-{\frac{4}{3}}s^2_W\ \ , \ \ d^{Z, SM0}_2=-g_{At}=-\,
 {\frac{1}{2}} \ \ . 
\end{equation}
SM at 1-loop and NP lead to
additional ($q^2$-dependent) contributions to these 
couplings, as well to the couplings of the second set, called set (2),
consisting of  
$d^{\gamma}_2(q^2)$, $d^{\gamma}_3(q^2)$ and $d^{Z}_3(q^2)$.\par

Departures from the SM (tree + 1-loop) are  defined as:
\bq 
\bar d^V_j \equiv d^V_j-d^{V,SM}_j \ \ \ . \ 
\label{eq:dvbar}
\eq
The
explicit expressions of the contributions to the $\bar d^V_j$ 
for each  $\O_i$ at tree
level and at 1-loop are given in  \cite{Kur}. 
We will now precisely determine the accuracy at
which they can be observed or constrained.\par

The reaction $e^+e^-\to t\bar t$ offers a way to determine all six
couplings by studying the top quark spin density matrix
elements. There are six independent such elements
$\rho^{L,R}_{++,--,+-}$, (where $L,R$ refer to the longitudinal 
polarization of the $e^-$ beam), which can be reconstructed
through the decay chains $t\to Wb \to l\nu b$ and  $t \to W b \to
q\bar q'b $, for any $q^2$ and production angle $\theta$ 
\cite{Kur, dj}.  Measurements of the magnitude
of density matrix elements (like the production cross section
$\sigma_{t\bar t}$) are affected by the lack of precise knowledge of the
$t\to Wb$ decay width \cite{treview, effic}. 
More accurate are the measurements of asymmetries 
like forward-backward or L-R asymmetries, which, being 
ratios of cross sections, are free
of this uncertainty. In order to study the 
$\gamma t\bar t$ and  $Z t\bar t$ couplings, 
we have tried to rely as much as 
possible on such observables.\par

In the general 6-parameter case this is not fully possible.
Asymmetries only depend on two combinations of 
the first set of the three $\bar d_j^V$ couplings. 
To get a third one, information
sensitive to the absolute normalization, like
$\sigma_{t\bar t}$, is required (which also depends on the top quark decay
width). For the second set of couplings we do not have this
problem, since asymmetries are now sensitive to
ratios of the type $\bar d/d^{SM}$ and allow to constrain
the complete set (2). We have made applications to an LC collider 
\cite {LC} with $0.5$,
$1$ and $2$ TeV and luminosity of $20$, $80$ and $320$ $fb^{-1}$
respectively, leading to more than $10^{4}$ events. A reconstruction
and detection efficiency of $18 \%$ \cite{effic} 
has been applied before
computing the statistical accuracy for each
observable. From this we have obtained the accuracy expected in the
determination of the top quark density matrix elements and the
constraints on the NP couplings $\bar d_j$ (ellipsoid in 6-parameter
space). Examples are shown in Fig.1a,b. The complete set of results
is given in \cite{dj}.\par

The essential features are the following.
For couplings of set (1), a band of width $\pm0.02$ 
appears, as information on the overall magnitude
of the amplitudes is missing when only asymmetries are used. 
When cross section
measurements are added assuming an uncertainty of $20\%$ ($2\%$) 
on the top decay width, a constraint of the order of 
$\pm0.05$ ($\pm0.02$)  arises. On the contrary, for couplings of set
(2), asymmetries alone constrain the couplings at the level of 
$\pm0.02$ in the
unpolarized case and at the level of $\pm0.01$ in the polarized 
one.\par

We have also studied constrained cases like the 4-parameter one
obtained  by restricting to the NP forms generated by 
the operators $\O_{t2}$, $\O_{Dt}$, 
$\O_{tW\Phi}$ and $\O_{tB\Phi}$ contributing to the $\gamma t\bar
t$ and $Z t\bar t$ vertices at tree level \cite{Kur}. 
This case is further reduced to a three parameter one if only
the operators $\O_{t2}$,  
$\O_{tW\Phi}$ and $\O_{tB\Phi}$ generated in the dynamical
models studied in \cite{Papa} are retained. 
Finally, a two parameter case is met  if one insists on a chiral 
form for NP where only anomalous left-handed and
right-handed $Z t\bar t$ couplings appear \cite{tchiral}.\par

In these constrained cases it is sufficient to use only
asymmetries, and the typical accuracy 
for both sets of couplings is of the order of $\pm0.02$ in the
no polarization case,  and of the order of 
 $\pm0.01$ when polarization is available.
Finally, the results for all operators listed in Class 1
\cite{Papa}, can be found in Table 7 of \cite{dj}. 
Essentially there are two levels of constraints. For the four operators 
$\O_{t2}$, $\O_{Dt}$, $\O_{tW\Phi}$ and
$\O_{tB\Phi}$ contributing at tree level at the LC,
the constraints reach effective scales lying in the 5 to
50 TeV range, which is much larger then the range  
reached through the  indirect result arising from the 
Z-peak contributions. On the contrary for the operators $\O_{qt}$,
$\O^{(8)}_{qt}$, $\O_{tb}$ contributing at loop level to $e^-e^+
\to t \bar t$, the Z-peak
constraints are not expected to be substantially improved, and
the NP scales should  lie in the 1-5 TeV range. 
However for the two  operators
$\O_{tt}$ and $\O_{tG\Phi}$, which contribute only at loop level 
and do not give substantial effects   at Z-peak, 
interesting constraints in the 5-10 TeV range will arise 
from $e^-e^+\to t \bar t$.  \par

\subsection{Tests in $e^+e^-\to b\bar b$}\par
We now discuss the $e^+e^-\to b\bar b$ channel. For  the
operators in Class 1 and Class 2 \cite{Papa}, the aforementioned channel
at LC cannot improve the existing information by other means.
Thus for the Class 2 operators which give tree level
contributions to the 
$\gamma b\bar b$ and $Zb\bar b$ couplings
\cite{Bing-Lin},  the $0.5\%$  accuracy obtained from
the $Z\to b\bar b$ measurements pushes the NP scale in the
10 TeV range; and this cannot be improved by new measurements in the
$e^+e^-\to b\bar b$ channel at higher energies. This is also true
for the operators of Class 1, which contribute only at loop level to the 
$\gamma b\bar b$ and $Zb\bar b$ vertices.
Some of these operators are in fact constrained by Z-peak
measurements, because their contribution there is  enhanced by
$m^2_t/M^2_W$ factors \cite{Zbbtop}. 
We have also seen that in some cases, like for 
 $\O_{qt}$, $\O^{(8)}_{qt}$ and $\O_{tb}$, Z peak measurements give 
better constraints than $e^+e^-\to t\bar
t$ at high energies. \par

Nevertheless, we have found that 
$e^+e^-\to b\bar b$ measurements beyond Z-peak can improve the constraints
for certain  "non-blind" Class 3 operators containing
derivatives of gauge fields \cite{Papa}.  
This turns out to be the case for the  gauge boson
operators  $\O_{DW}$, $\O_{DB}$ studied in \cite{Hag1,clean},
and the operators  $\O_{qW}$,  $\O_{qB}$, $\O_{bB}$ studied 
in \cite{bop}. 
They all give effective
four-fermion $e^+e^-b\bar b$ contact interactions,
whose contribution is enhanced with respect to
the Z, $\gamma$ ones, by  $q^2$ factors. This property
allows the disentangling of these contributions from those 
of all other operators, by using the so-called 
"Z-peak subtraction method" \cite{Zsub}.
This method  consists in using as inputs $\Gamma(Z\to f \bar f)$, $A_f$
instead of $G_{\mu}$, $s^2_W$. Thus, it automatically ~absorbs 
the  NP effects at Z peak and leaves only room for those ones which
can survive in the difference of structure functions of the type
\bq  
F_i(q^2)-F_i(M^2_Z) \ \ ,
\eq
which are combinations of self-energies, vertices, box and
NP contributions.
This way, only the contributions from the aforementioned 
Class 3 operators 
 $\overline{\O}_{DB}$, $\overline{\O}_{DW}$, $\O_{qW}$, 
$\O_{qB}$, $\O_{bB}$ survive sufficiently beyond the Z-peak.  
The operators  $\overline{\O}_{DB}$ 
 $\overline{\O}_{DW}$ are  constrained from $e^+e^-\to
l^+l^-$ and $e^+e^-\to q\bar q$ and have been studied in 
\cite{Hag1,clean}; while  $\O_{qW}$, 
$\O_{qB}$, $\O_{bB}$ can  be studied by measuring the 
observables $\sigma_{b}$,
$A^{b}_{FB}$, $A^b_{LR}$, $A^{pol(b)}_{FB}$,
which of course require polarized beams \cite{bop}. 
At LEP2 polarization is not available and
we only get two constraints for these three operators. 
This is the origin of
the band appearing in Fig.2; (illustrations for other couples of
operators can be found in ref.\cite{bop}).  
For polarized beams at LC, the system
can be completely constrained leading to sensitivity limits 
on NP scales in the range 30-50 TeV; indeed
a rather high level.\par

In conclusion, we summarize the  constraints 
obtained or expected from for the whole set of
$dim=6$  operators. We give below the range of NP scales 
(in TeV) up to which experiments at LEP and LC can be
sensitive.\\ 
\begin{tabular}{ccc}
\multicolumn{3}{c}{ Bosonic operators} \\
non-blind &   17~ (LEP2) &   50 ~(LC)\\
blind (TGC) &    1.5 ~(LEP2) &  20 ~(LC)\\
superblind(Higgs)  &  7 ~(LEP2) &  20 ~(LC), 
70 (LC$\gamma\gamma$)\\ \\
\multicolumn{3}{c}{Heavy quark operators } \\
Class 1 &    5 ~(LEP1) &     10-50 ~(LC)\\
Class 2 &   10 ~(LEP1) &    10 ~(LC)\\
Class 3 &   8 ~(LEP2) &     30-50 ~(LC)
\end{tabular}

\vspace*{.5cm}

As one can see, limits expected from the LC should improve by
one order of magnitude the ones indirectly obtained at LEP1
or expected from LEP2. They should lie in the several tens of TeV
range. This is
a domain which covers various types of models beyond SM (Technicolour,
extended gauges,...). Our hope is that experiments 
in various sectors will
reveal certain correlations which could
select a few operators of our list and give  hints for the structure
of new physics.\par

\section {Tests  of CP violating $H\gamma \gamma$ interactions}\par

The possibility to have polarized $e^-,~e^+$ beams in a 
linear collider, renders them  very efficient in
searching for CP violation among the  NP
interactions. For example, by measuring $e^-e^+ \to W^-W^+$ 
using polarized  beams, it is possible to improve considerably
\cite{Papadopoulos}, the present constrains on 
the CP violating triple gauge boson interactions \cite{DeR-CP}. 
Moreover at a LC, it is possible to 
search for CP violating forces directly connected to the Higgs
particle \cite{Stong, Kramer, Higgs-R}.\par

Here we concentrate on the option to use LC as a 
$\gamma \gamma$ collider by backscattering polarized laser
photons from
polarized electrons and positrons \cite{laser}. In particular we
concentrate on studying the CP violating forces affecting 
the $\gamma \gamma \to H$ process. In the philosophy followed
in the present note, where all new physical degrees of freedom 
are too heavy to be directly produced, such CP violation  
may arise from the $dim=6$ operators called $\tilde{\O}_{WW}$,
$\tilde{\O}_{BB}$ in \cite{Stong, Choi, Tsirigoti-CP},
whose couplings we denote by $\bar d$ and $\bar d_B$
respectively. \par

As soon as the Higgs particle is discovered,
a general study of all Higgs properties, including of course
the possibility for CP violating forces, will become mandatory.
For CP violation in particular, such studies
 are based on constructing appropriate spin asymmetries
with respect to the polarizations of the laser photons and 
$e^-, ~e^+$ beams.  For Higgs production through $\gamma \gamma$
collisions these are constructed by noting that for a laser
photon described by the density matrix
\bq
\label{laser-rho}
\rho_{laser}^N ~ = ~\frac{1}{2}
\left (\matrix{1+P_\gamma & -P_t e^{-2i\varphi} \cr
-P_t e^{+2i\varphi} & 1-P_\gamma } \right ) \ \  
\eq
in its helicity basis, the corresponding photon obtained by
backscattering the laser one from the $e^-$ beam is described 
by the density matrix \cite{laser}
\bqa
\label{Back-rho}
\rho^{BN} & =& \frac{1}{2}
\left (\matrix{1+\xi_2(x) & - \xi_{13}(x) e^{-2i\varphi} \cr
- \xi_{13}(x)  e^{+2i\varphi} & 1-\xi_2(x) } \right ) \ \ ,
\eqa 
where $P_\gamma$ describe the circular polarization of the laser
photon, $(P_t, ~\varphi)$ give the magnitude of the linear
polarization and its azimuthal angle around the momentum 
of laser-photon, while the Stokes parameters
$\xi_j$ of the backscattered
photon are known functions of $P_\gamma, ~P_t,~\varphi$
and of the (longitudinal) polarization $P_e$ of the 
$e^-$ beam \cite{laser}. Corresponding quantities (denoted as
barred below) are analogously  defined for the photon 
backscattered from the positron also.\par

The two possible asymmetries are ($\chi\equiv \varphi -
\bar\varphi$)  
\bq
\label{Alintilde}
\tilde A_{lin} ~ = ~\frac{|
N_{\tau_H}(\chi=\frac{\pi}{4})-N_{\tau_H}(\chi=-
\frac{\pi}{4})|}{N_{\tau_H}(\chi=\frac{\pi}{4})
+N_{\tau_H}(\chi=- \frac{\pi}{4})} ~ = ~
\frac{\langle \xi_{13}\bar \xi_{13}\rangle }
{1+\langle \xi_{2}\bar \xi_{2}\rangle }~ A_{lin} \ \ 
\eq
which uses the relative angles of the linear polarizations of
the two photons, and 
\bq
\label{Acirctilde}
\tilde A_{circ} ~=~ \frac{|N_{\tau_H}^{++}-N_{\tau_H}^{--}|}
{N_{\tau_H}^{++}+ N_{\tau_{H}}^{--}}~ =~
\frac{\vert \langle \xi_2+\bar \xi_2\rangle \vert}
{1+\langle \xi_{2}\bar 
\xi_{2}\rangle }~ A_{circ} \ \ ,
\eq
where the signs in the upper indices describe generically two
opposite signs for the pairs $(P_e, ~P_\gamma)$ and $(\bar P_e,
~\bar P_\gamma)$, which generates an asymmetry between 
two opposite circular polarizations
for the photons backscattered from $e^-$ and $e^+$
respectively.\par 

The quantities $A_{lin}$ and $A_{circ}$ defined in 
(\ref{Alintilde} ,\ref{Acirctilde}), are sensitive to the combination
$\bar d \swd +\bar d_b \cwd$ of the NP couplings, and depend of
course also on the SM dynamics.  It turns out that
$A_{circ}$ is useful only for $200GeV \lsim \mh $; while 
$A_{lin}$ can be  useful for both $100 GeV
\lsim \mh \lsim 150 GeV$ as well as for $200GeV \lsim \mh $
\cite{Tsirigoti-CP}. The reachable sensitivity limits 
are indicated in Tables I, II 
(where $x_0\equiv 4E\omega_0/m_e^2\leq 2(1+\sqrt{2})$ 
is the usual parameter
determined from the laser energy) and background effects have to
some extent been taken into account \cite{Tsirigoti-CP}.\par

It is  concluded from this that using polarized beams for realizing 
the $\gamma \gamma$ colliders, it is possible to
construct observables sensitive only to the 
CP violating NP couplings and thereby distinguish
them  from the CP conserving
ones. This is not attainable for  unpolarized beams.
Thus, the process 
$\gamma \gamma \to H$ at a $0.5TeV$ tunable linear collider may be
able to put  limits on the NP coupling  
$\bar d \swd +\bar d_B \cwd$ at the $10^{-3}-10^{-4}$ level, 
for Higgs masses in the ranges
$100 \lsim \mh \lsim 150 GeV$ and $200 \lsim \mh \lsim 350 GeV$.
This way, one can probe NP scales in the range of 10-20TeV
for NP generated by $\tilde{\O}_{WW}$, 
and in the range of 30-50TeV in the case $\tilde{\O}_{BB}$.\par
This information is complementary, and at least an order of
magnitude  more precise than the one attainable through
production   of $WW$ pairs in $\gamma \gamma $ 
collisions, where the same combination of NP couplings
is measured \cite{Choi, Higgs-H, Higgs-R}. 
Information on independent combinations of the CP violating
couplings at the level of $10^{-2}$
may be obtained by looking at $e^-e^+ \to H\gamma,~HZ$
\cite{Vlachos, Stong, Kramer}. Thus, a combination of such
measurements should be able to constrain separately each of the two
CP violating couplings $\bar d$ and $\bar d_{B}$ at
the $10^{-2}$ level \cite{Tsirigoti-CP}.


\renewcommand{\arraystretch}{1.2}
\begin{table}
\begin{small}
\begin{tabular}{|c|c|c|c|} \hline
\multicolumn{4}{|l|}{{\bf TABLE I.} $3\sigma$ upper bounds on
CP-violating NP couplings from $A_{lin}$ asymmetry, using}\\ 
\multicolumn{4}{|c|}{$H \to b \bar b$ for the two polarization
choices:}\\
\multicolumn{4}{|c|}{$P_t=\bar P_t=1, ~P_\gamma=\bar P_\gamma =0$, 
 \hspace*{0.5cm} ($P_t=\bar P_t=P_\gamma=\bar P_\gamma= 1/\sqrt 2, 
~P_e=\bar P_e=1$).}\\
\hline\hline
\multicolumn{1}{|c|}{$m_{H}$ (GeV) } &
\multicolumn{1}{|c|}{$x_0$} & \multicolumn{1}{|c|}{upper limit
on ${\bar d_{B}} c_{W}^{2}+\bar d s_{W}^{2}$} &
\multicolumn{1}{|c|}{$(\varphi-\bar \varphi)$-averaged Events }\\ \hline
 & 0.5 & $10^{-3}$  (2x$10^{-3}$) & 197 (168) \\ \cline{2-4}
100 & 0.8 & 2x$10^{-2}$  (1.6x$10^{-2}$) & 184 (233) \\ \cline{2-4}
 & 1 & 0.12  (0.05) & 160 (255) \\ \hline
 & 0.5 & 5x$10^{-4}$  (8.5x$10^{-4}$) & 200 (160) \\ \cline{2-4}
 & 0.8 & 8x$10^{-3}$  (6x$10^{-3}$) & 207 (223) \\ \cline{2-4}
120 & 1 & 0.05  (0.017) & 192 (264) \\ \cline{2-4}
 & 1.5 & 0.16  (0.11) & 175 (306) \\ \hline
 & 0.5 & 4x$10^{-4}$  (6.5x$10^{-4}$) & 144 (111) \\ \cline{2-4}
 & 0.8 & 5x$10^{-3}$  (3.2x$10^{-3}$) & 159 (143) \\ \cline{2-4}
140 & 1 & 0.03  (0.01) & 156 (181) \\ \cline{2-4}
 & 1.5 & 0.09  (0.06) & 147 (237) \\ \cline{2-4}
 & 2 & 0.22  (0.22) & 138 (252) \\ \hline
 & 0.5 & 5x$10^{-4}$  (8x$10^{-4}$) & 85 (64) \\ \cline{2-4}
 & 0.8 & 4.3x$10^{-3}$  (3x$10^{-3}$) & 106 (87) \\ \cline{2-4}
150 & 1 & 0.023  (0.008) & 107 (113) \\ \cline{2-4}
 & 1.5 & 0.08  (0.05) & 102 (157) \\ \cline{2-4}
 & 2 & 0.2  (0.2) & 97 (174) \\ \hline 
\end{tabular}
\end{small}
\end{table}

\begin{table}
\begin{small}
\begin{tabular}{|c|c|c|c|} \hline
\multicolumn{4}{|l|}{{\bf TABLE II.} $3\sigma$ upper bounds on
CP-violating NP couplings from $A_{circ}$ asymmetry, using}\\
\multicolumn{4}{|c|}{$H\to ZZ \to l^-l^+X$ decay 
and circularly polarized laser beams with
$P_e=\bar P_e =-P_\gamma=-\bar P_\gamma=\pm 1 $.} \\
\hline\hline
\multicolumn{1}{|c|}{$m_{H}$ (GeV) } &
\multicolumn{1}{|c|}{$x_0$} & \multicolumn{1}{|c|}{upper limit
on ${\bar d_{B}} c_{W}^{2}+\bar d s_{W}^{2}$} &
\multicolumn{1}{|c|}{$\xi_2$-averaged Events}\\ \hline
 & 1 & 5.7x$10^{-4}$ & 141 \\ \cline{2-4}
 & 1.5 & 6x$10^{-4}$ & 133 \\ \cline{2-4}
200 & 2 & 6.2x$10^{-4}$ & 121 \\ \cline{2-4}
 & 2.5 & 6.5x$10^{-4}$ & 112 \\ \cline{2-4}
 & 4 & 7x$10^{-4}$ & 95 \\ \cline{2-4}
 & 4.82 & 7.3x$10^{-4}$ & 89 \\ \hline
 & 1 & 6.3x$10^{-4}$ & 49 \\ \cline{2-4}
 & 1.5 & 4.2x$10^{-4}$ & 109 \\ \cline{2-4}
240 & 2 & 4.3x$10^{-4}$ & 105 \\ \cline{2-4}
 & 2.5 & 4.4x$10^{-4}$ & 99 \\ \cline{2-4}
 & 4 & 4.7x$10^{-4}$ & 85 \\ \cline{2-4}
 & 4.82 & 5x$10^{-4}$ & 80 \\ \hline
 & 1.5 & 4.5x$10^{-4}$ & 55 \\ \cline{2-4}
 & 2 & 3.7x$10^{-4}$ & 78 \\ \cline{2-4}
280 & 2.5 & 3.7x$10^{-4}$ & 78 \\ \cline{2-4}
 & 4 & 4x$10^{-4}$ & 70 \\ \cline{2-4}
 & 4.82 & 4x$10^{-4}$ & 66 \\ \hline
 & 2 & 4.8x$10^{-4}$ & 31 \\ \cline{2-4}
320 & 2.5 & 3.6x$10^{-4}$ & 53 \\ \cline{2-4}
 & 4 & 3.5x$10^{-4}$ & 56 \\ \cline{2-4}
 & 4.82 & 3.6x$10^{-4}$ & 54 \\ \hline
 & 2.5 & 6x$10^{-4}$ & 15 \\ \cline{2-4}
350 & 4 & 3.4x$10^{-4}$ & 45 \\ \cline{2-4}
 & 4.82 & 3.4x$10^{-4}$ & 46 \\ \hline
\end{tabular}
\end{small}
\end{table}


\def\x{$\bar d^\gamma_1$}
\def\y{$\bar d^\gamma_2$}
\def\z{$\bar d^\gamma_3$}
\def\u{$\bar d^Z_1$}
\def\v{$\bar d^Z_2$}
\def\w{$\bar d^Z_3 $}

\begin{figure}[p]
\vspace*{1.cm}
\hspace{4cm} \u
\vspace*{-3cm}
\[ \hspace*{1cm}
\epsfig{file=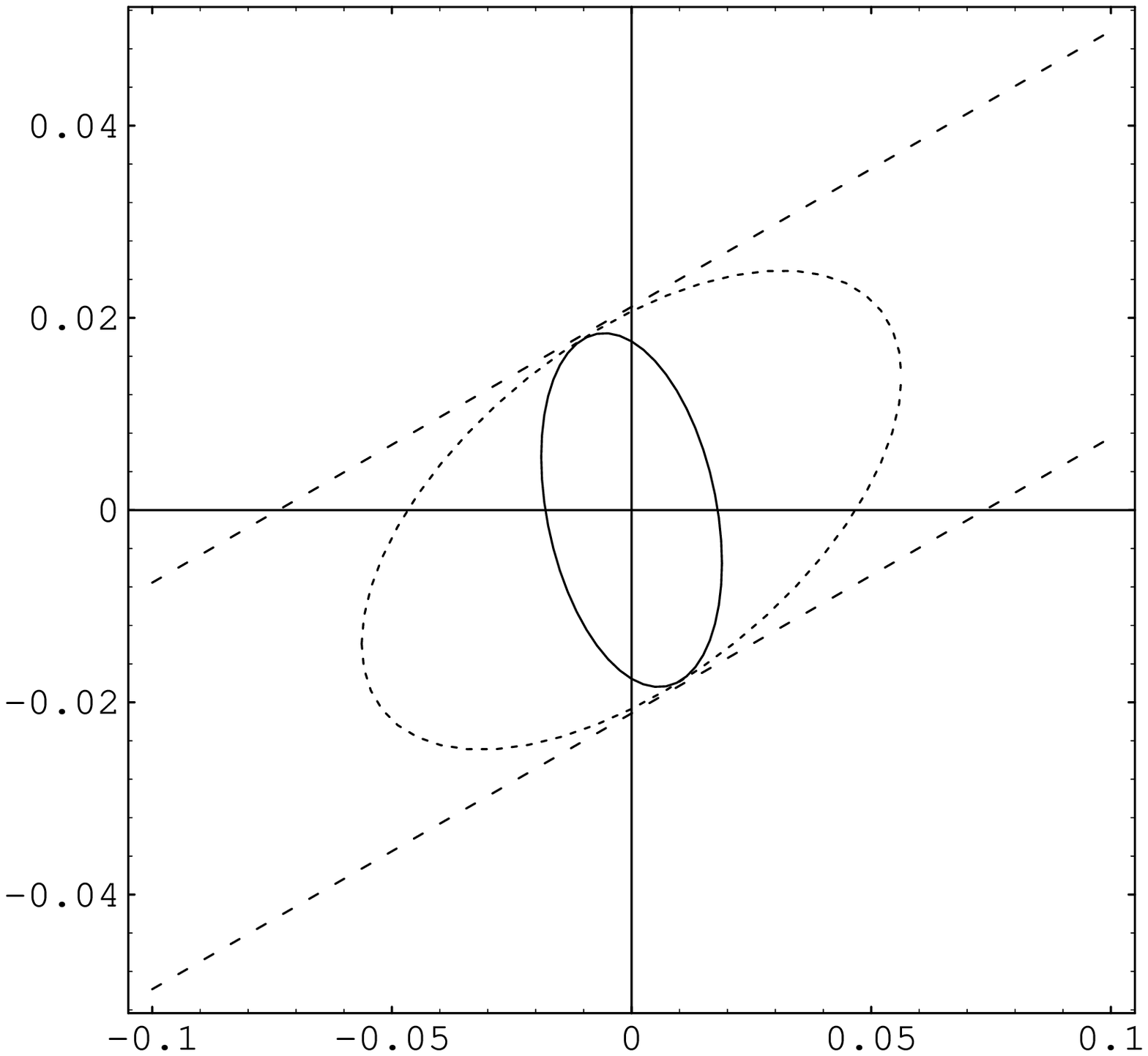,height=11cm}\]\\

\vspace*{-3.cm}
\hspace{11.5cm} \x

\vspace*{-0.2cm}
\hspace*{8cm} (a)\\

\vspace*{0.2cm}
\hspace{4cm} \w


\vspace*{-2.7cm}
\[\hspace*{1cm}
\epsfig{file=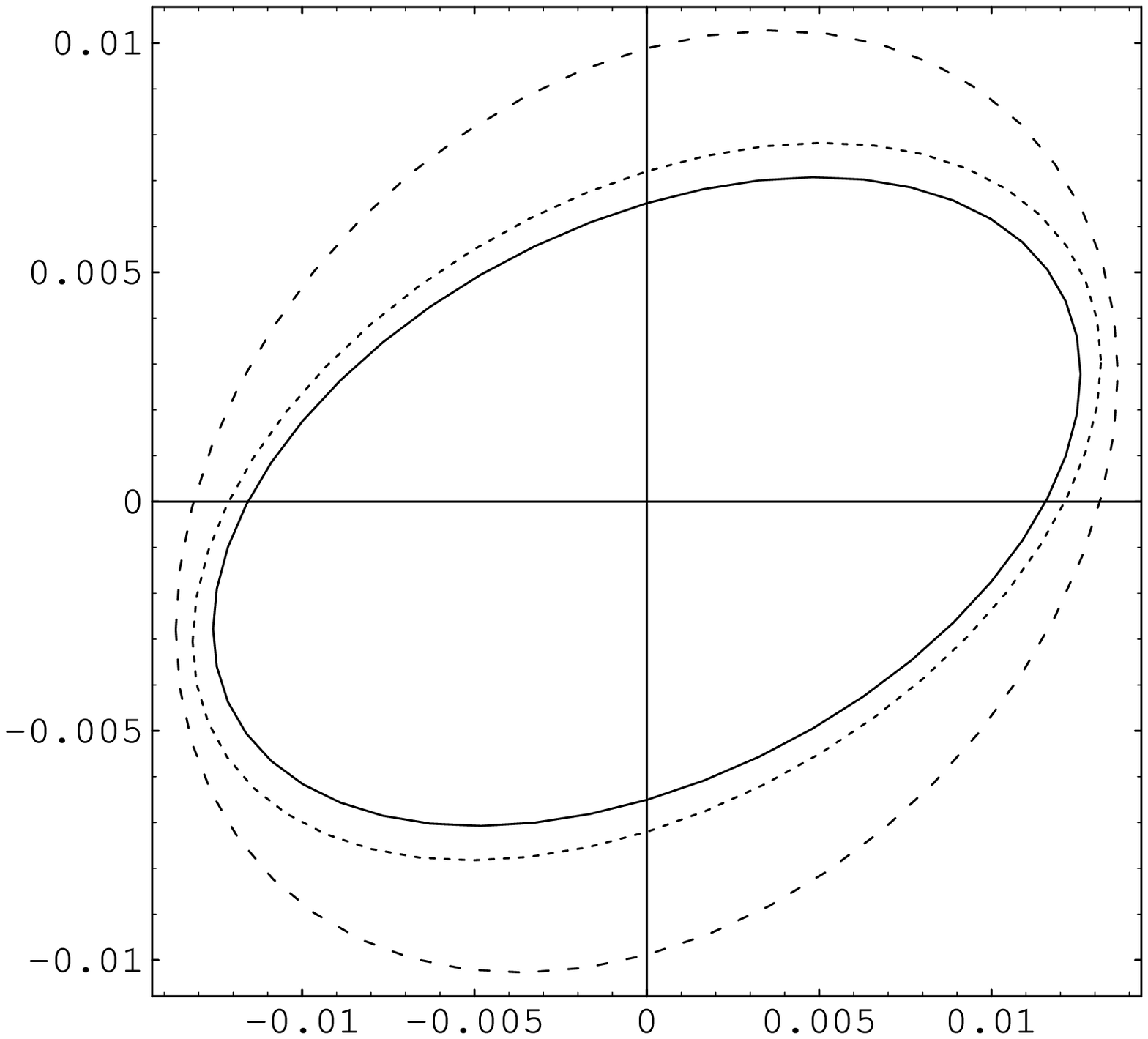,height=11cm}\]\\

\vspace*{-3cm}
\hspace*{11.5cm} \y\\
\vspace*{-.5cm}
\hspace*{8cm} (b)

\vspace*{0.5cm}
\begin{center} 
\caption[1]{Observability 
limits in the 6-parameter case;
(a) among couplings of set (1),\\ 
(b) among couplings of set (2);
from asymmetries alone (- - - -), from asymmetries
and integrated observables with a normalization uncertainty of
2\% (........), 20\% (----------).}
\end{center} 
\end{figure}

\begin{figure}[p]
\begin{center}
\mbox{
\epsfig{file=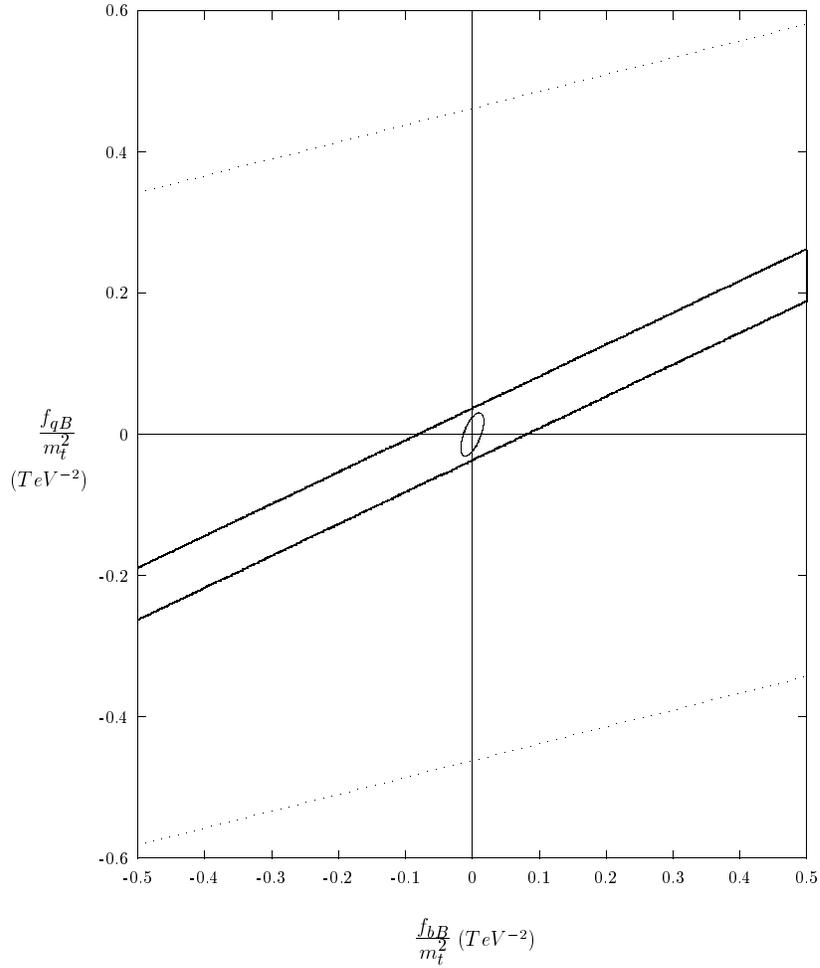,height=20cm}}
\vspace*{-2cm}
\caption[1]{ Constraints from $e^+e^-\to b\bar b$ 
observables in the
3-free parameter case, projected on the ($f_{qB}$, $f_{bB}$) plane; 
at LEP2 (without polarization) ({\it dotted}), ~at 
NLC (without polarization) ({\it solid}),
at NLC (with polarization) ({\it ellipse}).}
\end{center}
\end{figure}

\end{document}